\documentstyle[epsf,axodraw,amssymb]{article}
\begin{document}

\title{
      {\bf MATHEMATICAL PHYSICS AND LIFE}%
\footnote{To appear in Mathematical Sciences Series, Vol.4:
{\it Computing and Information Sciences: Recent Trends}, ed. J.C. Misra,
Narosa (2003) 270-293, {\tt quant-ph/0202022}.}
}
\author{Apoorva Patel\\
CTS and SERC, Indian Institute of Science, Bangalore-560012}
\date{\vspace*{-2mm}\small E-mail: adpatel@cts.iisc.ernet.in \vspace*{-2mm}}
\maketitle

\begin{abstract}
\noindent
It is a fascinating subject to explore how well we can understand the
processes of life on the basis of fundamental laws of physics. It is
emphasised that viewing biological processes as manipulation of information
extracts their essential features. This information processing can be
analysed using well-known methods of computer science. The lowest level
of biological information processing, involving DNA and proteins, is the
easiest one to link to physical properties. Physical underpinnings of the
genetic information that could have led to the universal language of 4
nucleotide bases and 20 amino acids are pointed out. Generalisations of
Boolean logic, especially features of quantum dynamics, play a crucial role.
\end{abstract}

\section{What is Life?}

Often life is characterised on the basis of its fundamental processes:
metabolism and reproduction. But perhaps it is easier to characterise what
life is not than to define what it is. Life is not a process of thermodynamic
equilibrium---all the metabolic activities are driven by a continuous supply
of energy which is obtained from the environment as food. Evolution of life
is not dictated by precise rules of logic---it is governed by random events
where memory plays a role but there is no foresight.  Living organisms are
made of neither completely regular arrangements as in a solid nor totally
random arrangements as in a gas---all the activities take place in a liquid
medium with continuous jostling amongst various biomolecules. These properties
make the familiar tools of physicists---equilibrium dynamics, axiomatic
deductions, periodic structures, perturbation theory---totally useless when
it comes to understanding life. Life is termed too nonlinear, too complex,
too unwieldy for the simple tools of physicists to handle.

Information theory provides a way out of this apparently hopeless conundrum.
Information is the abstract mathematical concept that quantifies the notion
of order amongst the building blocks of a message. It extracts just the right
features from the apparently chaotic ensemble that is life, and allows the
study of its manipulations without going into the nitty-gritty of all the
details. Of course, figuring out the appropriate information in a given
process requires detailed experiments and modeling. But once the essential
features have been extracted, the rules of information processing are precise
enough to allow their systematic analysis in a mathematical framework.

It can be argued that information processing is not just a particular
characteristic of living organisms, but it is their most important
characteristic \cite{schrodinger}.
The processes of life are aimed towards survival of the organism. Even
though we do not quite understand why the organisms want to perpetuate
themselves, we have enough evidence to show that they use all available
means for this purpose \cite{dawkins}.
The damage caused by the disturbances from the environment makes it
impossible for a particular individual to survive forever, so the
perpetuation is carried out through the process of replication. In this
process, the knowledge gathered on how to survive is carried forward from
one generation to the next. The pattern of acquiring, interpreting and
passing on information, often after using and refining it, has occurred
repeatedly during evolution. The hereditary genetic information is the
most basic and primitive communication system. As living organisms evolved,
the lower levels of information communication have been bypassed in favour
of more efficient higher level mechanisms, reducing dependence on direct
physical contact at every stage. Electrochemical signaling between cells
and nervous systems evolved in multicellular organisms, teaching and
imitation arose amongst parents and offsprings, verbal and sign languages
originated amongst members of the same species, humans created books and
libraries for long term storage of information, we use telecommunication
and computers nowadays, and the future (already imagined in science fiction
stories) is likely to witness fusion of brains with computers.

\section{Biological Information}

Living organisms absorb free energy from the environment to create order
from disorder. All their struggles for raw materials and energy are
ultimately for picking up necessary building blocks and organising them
in a desired manner. After the organisms die, the building blocks gradually
come apart and become random ensembles. Quantitatively the information
carried by a set of building blocks is nothing but the entropy (up to an
unimportant proportionality constant), and principles of computer science
can be used to study its transformations.

Table 1 lists various stages of information processing tasks carried out
by living organisms and compares them with similar tasks carried out by
our electronic computers. They have been arranged in a hierarchy of levels
according to the physical scales involved. Different levels typically have
different languages, and the processing translates the information from one
level to the next. We are accustomed to looking at our computers from the
top level down---from the abstract mathematical equations to the transistors
embedded in the silicon chips. On the other hand, living organisms have
evolved from the bottom level up---from the biomolecular interactions to
multicellular systems.

\begin{table}[tbh]
\vspace*{-2mm}
\begin{center}
\begin{tabular}{lll}
Living organisms         &  Task       & Computers                   \\
\hline
Signals from environment & Input       & Data                        \\
Sense organs             & High level  & Pre-processor               \\
Nervous system + Brain   & Translation & Operating system + Compiler \\
Electrochemical signals  & Low level   & Machine code                \\
Proteins                 & Execution   & Electrical signals          \\
DNA                      & Programme   & Programmer                  \\
\end{tabular}
\caption{Hierarchical processing of information in living organisms and
in computers.}
\end{center}
\vspace*{-4mm}
\end{table}

To understand the working of an information processing system, one must
figure out what happens at each level step by step. But where should one
begin, when little is known about the system? At this juncture, it is
worthwhile to observe that two major simplifications occur as one proceeds
from high level information processing to low level:\\
(1) The languages at the higher levels are abstract software. They get
translated from one to another according to established conventions. On
the contrary, the languages at the lowest levels are directly connected
to the physical responses of the hardware. There is no more translation
of the message---the interpretation of the signal is built into the design
of the system and not left to an external agency. (e.g. we can programme
the mathematical formulae into the computer in a variety of notations,
but ultimately they are all converted to pulses of voltages and currents
because the transistors in the silicon chips cannot respond to any other
language.)\\
(2) The abstract high levels allow a variety of subjective choices in the
languages. That leads to lots of variations, historical adaptations and a
large number of possible instructions. At the lowest level, only a handful
of instructions related to physical responses of the hardware are possible,
and the language tends to become universal. (e.g. for writing the computer
programmes, we may choose between Fortran and C, or Unix and Windows, but
finally all that is reduced to the operations of Boolean logic.)

These properties make it obvious that it is easiest to decipher an information
processing system at the lowest level, where the physical properties of the
hardware dictate the language and the instructions \cite{levels}.
Complex systems are then generated by putting together a large number of
simple ingredients. With the progressive miniaturisation of the silicon
chips, the lowest level of information processing in our computers nowadays
is at the scale of a micron. The lowest level of information processing for
the living organisms is at the molecular scale---a scale smaller by a factor
of thousand. The molecules involved there are DNA and proteins, and their
language is universal---the same all the way from viruses and bacteria
to human beings. So our first objective is to identify the information
processing tasks carried out by DNA and proteins, and then study the
relation between these tasks and physical properties of the molecules.
It is important to note that biological systems are not general purpose
computers, where the same CPU handles all the tasks. Living organisms have
evolved specialised components to perform specific tasks, which makes it
easy to identify the task of a component and study its implementation in
detail.

DNA is the read-only-memory of the living organisms. It is a linear chain
constructed from an alphabet of 4 nucleotide bases. Most of the time, the
information carried by it remains nicely protected in the double helical
structure. Whenever necessary, the information is ``copied'' into another
DNA molecule during replication and mRNA molecule during transcription.
The ``copying'' is not literal, rather it follows rules of complementary
base-pairing. The ``copy'' is assembled on the master template, by picking
up desired nucleotide bases one by one from the surroundings and joining
them together in a chain. The available nucleotide bases are floating around
randomly in the cell, and the base-pairing rules decide which one is the
correct one to insert at a specific location. Thus the only task carried out
by DNA is the sequential assembly of a chain of nucleotide bases, selected
from a random ensemble.

Proteins carry out almost all the processes required by the living cell, by
binding to various molecules. The binding is highly specific, very much like
a lock and key mechanism. For this purpose, proteins contain structurally
stable features, precisely located on the surface. Each protein has its own
unique shape and participates in its own unique process. Proteins are made
of one or more tightly folded polypeptide chains (they may contain other
ingredients but the role of these other ingredients is essentially chemical
and not structural). Polypeptide chains are synthesised from an alphabet of
20 amino acids, and other ingredients get added afterwards. Each polypeptide
chain is assembled on the mRNA template during translation when successive
segments of 3 nucleotide bases are mapped onto individual amino acids. Every
chain subsequently folds up into a three-dimensional structure, uniquely
determined by its amino acid sequence. The proteins are thus involved in two
information processing tasks. One is similar to that of DNA, i.e. sequential
assembly of a chain of amino acids selected from a random ensemble. The
other is to create a multitude of three-dimensional structures by folding up
linear polypeptide chains.

These tasks are carried out at the molecular scale. We know the physical
laws applicable there---classical dynamics is relevant, but atomic structure
and quantum dynamics cannot be ignored. We can now pose the question:
were we to design a processor to carry out the tasks of DNA and proteins,
what design would we come up with knowing all that we do about the physical
laws? The rest of the article investigates this question, and compares the
results to what is found in nature.

\section{Survival of the Fittest}

It is advantageous to process the information efficiently, and not in any
haphazard manner. The first attempt may not provide the best solution to
a problem, but the attempts do not stop there. Technological developments
are driven as much by new inventions as by untiring attempts to improve and
optimise earlier solutions. In general, information processing is optimised
following two guidelines: minimisation of physical resources (time, space,
energy etc.) and minimisation of errors. These guidelines often impose
conflicting demands on the solution, and specific trade-offs are made
depending on the details of the problem.

In case of living organisms, the optimisation criteria are paraphrased as
``Darwinian evolution''. The environment and competition drive the living
organisms to adapt to them by exerting selection pressures, and we can
illustrate that by many examples. We now understand the genetic basis behind
Darwinian evolution. No information processing system can be perfect, and
occasional errors in genetic information processing produce random mutations
of living organisms. The error rate is bound to be small in any reasonably
stable system, and mutations are small local fluctuations in DNA molecules.
The mutated organism is in essentially the same environment as the original
one, and both have to compete for the available resources. If the mutation
improves the ability of the organism to survive, the mutated organism grows
in number, otherwise it fades away. To what extent may all this be quantified?

Let the index $i$ label a set of coexisting species in a given environment,
and let $\phi_i(t)$ denote their populations at time $t$. The simplest
evolution scenario is where the future population of each species depends
linearly on the present populations of all the species,
\begin{equation}
\phi_i(t+1) = \sum_j M_{ij} \phi_j(t) ~,~~ \phi_i(t) \ge 0 ~.
\end{equation}
The diagonal terms $M_{ii}$ represent the individual rates of growth, while
the off-diagonal terms $M_{i\ne j}$ represent interactions between species.
Every species consumes resources from the environment; when the available
resources are limited, the total population cannot exceed a certain value
(a normalisation convention can be chosen such that a unit of each species
utilises the same amount of resources). Evolution gets really competitive
when the total population reaches its maximum value. In this stage,
\begin{equation}
\sum_i \phi_i(t) = N ~,~~ \sum_i M_{ij} = 1 ~.
\end{equation}
Evolutionary models obeying Eqs.(1,2) have often been used together with the
constraint $0 \le M_{ij} \le 1$. They describe Markovian evolution in the
language of classical probability theory. With the constraints on $M_{ij}$,
one can prove many inequalities and convergence properties. This is not the
interesting part of evolution, however.

Fixed amount of available resources correspond to conservation laws. The
most general evolution in such circumstances is described by ``orthogonal
transformations'' \cite{norm}. The generators for these transformations are 
antisymmetric matrices. Thus in a general evolutionary setting, it is more
appropriate to let $M_{ij}$ take positive as well as negative values. As a
matter of fact, situations of both positive and negative feedback occur
routinely in biological systems (e.g. catalysis and inhibition, symbiosis
and parasitic behaviour, defence mechanisms and cancer, etc.). When the
resources are limited, one gains only at the expense of someone else---the
formal setting is called ``zero-sum games''.

We can now look at how evolution changes, when the range of $M_{ij}$ is
extended to include negative values. Clearly extending the range of $M_{ij}$
cannot make the evolutionary algorithm any less efficient. On the contrary,
recent developments in quantum computation \cite{nielsen}
offer a hint. Quantum algorithms exploit two features to beat their classical
counterparts, superposition of states and cleverly designed destructive
interference. Superposition means letting all the states be in the same
place at the same time, and it can reduce the spatial resources required
for the algorithm exponentially. But superposition is not an option available
to coexisting biological species.  Even in absence of superposition,
destructive interference can be used to eliminate undesired states and to
reach the output state more quickly. The execution time is reduced at least
by a constant factor if not polynomially. Biological evolution occurs over
long time scales, and even a tiny change in the rate of growth is important
because it can translate into exponential changes in populations over a long
time. Negative values of $M_{ij}$ can indeed be interpreted as effects of
destructive interference, which help the species reach their asymptotic
populations faster. The lessons learnt from quantum computation then tell us
that ``orthogonal transformations'' offer a quicker way of picking a winner
amongst the contenders. These arguments are not a rigorous derivation, but
they allow us to infer that competition beats altruism and Darwinian
selection is a consequence of limited availability of resources \cite{zerosum}.

\smallskip\noindent{\bf Exercise:}
{\small Study evolution algorithms governed by Eqs.(1,2). Estimate the
speed-up obtained when $M_{ij}$ are allowed to become negative compared
to when they are restricted to be positive.}
\smallskip

Having justified that optimisation principles do play a role in biological
systems, let us analyse what is optimised and how. In designing more and
more efficient computers, we have explored the criteria mentioned at the
beginning of this section. To optimise spatial resources, we need elementary
hardware components that are simple and easily available, and yet versatile
enough to be connected together in many different ways. This is the typical
choice made at the lowest level of information processing, and complicated
systems are then constructed by packing a large number of components in a
small volume. It helps to have a small number of elementary components,
because that reduces the number of possible instructions and the connections
amongst the components. With a small instruction set, individual steps can
be implemented quickly and the time required for processing information is
cut down. In addition to these, in the implementation of any given task,
we have to find a software algorithm which requires smallest number of
components and smallest number of execution steps.

For our computers, it is also necessary to minimise the energy consumption
during processing. But this feature is surprisingly absent in biological
systems. The reason is that the elementary components of biological systems
are so simple and cheap, that they carry out their tasks with very little
energy---a single biomolecular interaction requires a million times less
energy than a Boolean logic operation with modern silicon transistors.
As a result, biological systems often exhibit a wasteful feature. Millions
of eggs and pollen grains are produced when a few would have sufficed to
propagate the species in a secure environment. Such overkills strengthen
the competition and enforce survival of the fittest.

\smallskip\noindent{\bf Exercise:}
{\small Observe that biological systems have evolved excellent amplifiers
for processing occurring at the molecular scale. Eyes and chlorophyll can
detect a few photons, smells can be identified with a few molecules and
sound amplitudes with a fraction of atomic size can be heard. Body 
movements almost always use levers in the configuration of mechanical
disadvantage; it is somewhat ironic that mechanical advantage corresponds
to gain in power which is the reciprocal of the gain in amplification.
Obtain quantitative data for such amplification processes.}
\smallskip

Faithful information processing requires that the error rate must be kept
in control. One strategy is to shield the system from unwanted external
disturbances, but that does not protect against internal fluctuations.
Internal errors are minimised by selecting an information processing
language based on discrete variables (as opposed to continuous variables).
Allowed values of fundamental physical variables are often continuous,
in which case a set of non-overlapping intervals of values can be chosen
as the discrete variables. This is the common procedure of digitisation.
The advantage is that the discrete variables remain unaffected, even when
the underlying continuous variables drift, as long as the drifts keep the
values within the assigned intervals. As a matter of fact, digitisation
allows elimination of bounded errors. When the discrete variables are
chosen as far apart from each other as possible in a given range of values,
misidentification is minimised and one has the largest protection against
errors. Even with digitisation, errors from large fluctuations cannot be
eliminated. It is a curious fact that biological systems take advantage of
a tiny error rate---with too many errors the organism will not be able to
survive, but without mutations there will be no evolution.

\section{Choice of Language}

We are now in a position to look at some examples of information processing
systems, and understand how well they implement the optimisation principles. 
Table 2 lists some of the languages used by living organisms, their basic
components and the operations performed on those components. Messages are
constructed by linking the basic components---the building blocks of the
language---in a variety of arrangements. The information contained in a
message depends on the values and locations of the building blocks. Any
language that communicates non-trivial information must have the flexibility
to arrange its building blocks in different ways to represent different
messages.

This notion is quantified by saying that messages are aperiodic chains of
the building blocks, and the information contained in a message is its
entropy, i.e. a measure of the number of possible forms the message could
have taken \cite{shannon}.
This definition tells us that the information content of a message can be
increased by eliminating correlations from it and making it more random.
It also tells us that local errors in a message can be corrected by building
long range correlations into it. But it does not tell us what building blocks
are appropriate for a particular message. The choice of building blocks
depends on the type of information and not on the amount of information.

\begin{table}[tbh]
\vspace*{-2mm}
\begin{center}
\begin{tabular}{llll}
Physical system & Information & Building blocks   & Operations               \\
\hline
Computer   & Numerical        & Digits    & $Z_n$ (modulo-$n$ arithmetic)    \\
Verbal languages & Positional & Syllables & $S_n$ (permutations)             \\
Nervous system   & Temporal   & Electrical pulses & $Z_2$ (on/off)           \\
Genes      & Positional       & Nucleotide bases  & Comparison (replication) \\
Proteins   & Structural & Amino acids & $Z \otimes T$ (translation, rotation)\\
\end{tabular}
\caption{Different types of information systems and their properties.}
\end{center}
\vspace*{-4mm}
\end{table}

A striking feature of all the types of information listed above is that they
all have digital form. The simplicity of instructions and the control over
errors offered by digitisation are too important to ignore. The fact that
digitisation necessarily approximates values of continuous variables does
not matter in practical applications. The reason is that no practical result
needs infinite precision; as long as the results are obtained within
predefined but non-zero tolerance limits they are useful \cite{bounds}.
The outstanding optimisation question then is to figure out the best way of
digitising a message, i.e. what should be selected as the building blocks
of the aperiodic chain.

When the languages are versatile enough, information can be translated from
one language into another by replacing one set of building blocks by another.
Nonetheless, physical principles are involved in selecting different building
blocks for different information processing tasks. For example, our electronic
computers compute using electrical signals but store the results on the disk
using magnetic signals; the former encoding is suitable for fast processing
while the latter is suitable for long term storage. It is therefore the
available hardware and the job to be carried out which dictate the optimal
building blocks in any information processing task.

Languages of biological systems have arisen by trial and error exploration,
and by accumulation of small modifications, and not by mathematical deduction
of the optimal choice. Historical baggage is often carried forward in such
situations; one gets stuck in local optima without discovering the globally
optimal features. The optimal solution can be reached only when there is a
long enough time for exploration \cite{english}.
Biologists have often viewed the languages of genes and proteins as ``frozen
accident''---it is such a vital part of life that any change in it would be
highly deleterious \cite{frozen}.
On the other hand, these languages are universal across all living organisms
and exhibit efficient packing of information (little correlation or close to
maximal entropy), which suggest that the optimal form may have been reached.
Keeping all these ideas at the back of our mind, let us step by step search
for the optimal way of implementing the tasks of DNA and proteins, and
compare that to what is observed in nature.

\smallskip\noindent{\bf Exercise:}
{\small Understand the physical reasons behind various numerical systems we
use. Why do we have binary arithmetic for computers, decimal system for
counting, and even more complicated system for marking time?}

\section{Structure of Proteins}

It is convenient to analyse the language of proteins first, since it is a
straightforward exercise in classical geometry \cite{carbon}.
The problem is to find a discrete language than can encode arbitrary shapes
in 3-dimensional space, much like designing children's playing blocks so
that they can be linked together to create many different structures. This
means discretising the continuous operations of translation and rotation.
At the molecular scale, the atomic structure of matter automatically provides
such a discretisation of space, i.e. a lattice. Among all possible lattice
discretisations, the simplest lattice unit that can faithfully describe the
space in any dimension is the simplex---the volume element formed by a set of
$(d+1)$ points in $d$-dimensions \cite{cartesian}.

\begin{table}[tbh]
\vspace*{-2mm}
\begin{center}
\begin{tabular}{lll}
                & 1-dim          & 3-dim       \\
\hline
Information     & Numerical      & Structural  \\
Discrete space  & Integers       & Lattice     \\
Basic variables & \{0,1\}        & $l$=\{0,1\} \\
Implementation  & On/Off         & Tetrahedron \\
Operations      & Addition       & Translation \\
                & Multiplication & Rotation    \\
\end{tabular}
\caption{Comparison of the simplest 1-dim and 3-dim information encodings.}
\end{center}
\vspace*{-4mm}
\end{table}

The changes required to go from the familiar language of 1-dimensional
numerical information to the language of 3-dimensional structural information
are summarised in Table 3. We can quickly observe the following facts:\\
$\bullet$ In 3-dimensional space, the simplex is a tetrahedron. One can go
from any vertex to any other vertex by a single step. Arbitrary structures
can be constructed by joining tetrahedra together.\\
$\bullet$ A regular tetrahedron describes four equivalent and equidistant
states. They correspond to the smallest two representations of the spherical
harmonics, $l=0,1$. This is the obvious generalisation of the 1-dimensional
binary arithmetic to the 3-dimensional space. In quantum theory, these four
states are formed by $sp^3$-hybridisation of atomic orbitals and provide an
orthogonal basis.\\
$\bullet$ The ideal chemical element for realising tetrahedral geometry in
molecular structures is carbon, in the form of a diamond lattice. Carbon has
the capability to form aperiodic chains, where different side chemical groups
hang onto a backbone, which is a must for encoding information.\\
$\bullet$ The carbon atom is located at the centre of the tetrahedron, with
its four bonds directed towards the four vertices. The tetrahedral symmetry
group $T_d$ has 24 elements, which can be factored into a group $T$ of 12
proper rotations and reflection (or parity). The proper rotations consist
of rotations around four 3-fold axes and three 2-fold axes, and describe
rotations around bonds of the central carbon atom. The parity transformation
flips the chirality of the molecule, and is rare enough to be ignored in the
structural analysis of proteins.\\
$\bullet$ The tetrahedral geometry also plays a crucial role in the properties
of water---the essential solvent for all the processes of life. The proteins
are often hydrated, with water molecules filling the cavities. The role of
water, however, will not be pursued here.

Arbitrary irregular shapes can be constructed by gluing tetrahedra together,
similar to how children assemble their playing blocks. But such a process has
to rely on an external agency to stop the growing process when the desired
structure has been reached. At the lowest level of information processing,
there is no luxury of an external agency---the building blocks themselves have
to carry the information about the required shape. The solution is to encode
the 3-dimensional structural information as a 1-dimensional chain that knows
how to bend and fold at every step. (The assembly problem, which cannot do
without an external agency, then simplifies---only 1-dimensional chains have
to be assembled, and we will look at that later.) This is the solution chosen
by nature. The other advantage of constructing proteins as chains which can
fold and unfold is that they can cross membranes and cell walls, which they
need to do to participate in various cellular processes, without making big
holes in them. It is a marvelous piece of engineering that proteins unfold
to their chain form, cross the membrane through a small hole that prevents
other objects from leaking, and then fold again into their characteristic
shapes.

\subsection{The polypeptide backbone}

To have the capability to fold and unfold the backbone of the chain must
be strong and its side interactions weak. The simplest backbone formed by a
string of carbon atoms is that of polyethylene (i.e. $(-CH_2-)_n$). It is
made of rotatable carbon-carbon single bonds, and is too flexible to hold
its shape. The rigidity of the backbone can be increased by including in
it some double bonds that cannot rotate; the backbone will then have stiff
segments alternating with flexible joints similar to a chain of metal rings.
The simplest such non-planar backbone is formed by alternating one double bond
with two single bonds. That is the backbone of polypeptide chains---the stiff
$C-N$ peptide bond increases its rigidity, while the rotatable $C_\alpha$
bonds permit construction of a variety of 3-dimensional structures. The
polypeptide chain is synthesised by joining together amino acids, and their
chemical structures are shown in Fig.1. Every amino acid contains an acidic
$-COOH$ group and a basic $-NH_2$ group, and the peptide bond is formed by
their acid-base neutralisation. The amino acids differ from each other by
their R-groups; these R-groups hang on the sides of the polypeptide
backbone and interactions amongst them decide how the backbone would bend
and fold.

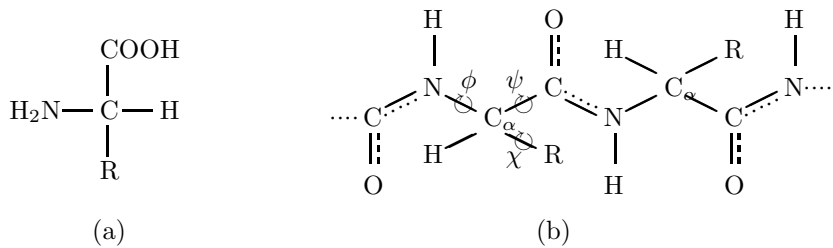
\begin{figure}[tbh]
{
\setlength{\unitlength}{1mm}
\begin{picture}(120,36)
  \thicklines
  \put(21,3){\makebox(0,0)[bl]{(a)}}
  \put(22,19){\makebox(0,0)[bl]{C}}
  \put(21,20){\line(-1, 0){4}}
  \put(10,19){\makebox(0,0)[bl]{H$_2$N}}
  \put(25,20){\line( 1, 0){4}}
  \put(30,19){\makebox(0,0)[bl]{H}}
  \put(23,22){\line( 0, 1){4}}
  \put(22,27){\makebox(0,0)[bl]{COOH}}
  \put(23,18){\line( 0,-1){4}}
  \put(22,11){\makebox(0,0)[bl]{R}}

  \put(80,3){\makebox(0,0)[bl]{(b)}}
  \put(53,19){\circle*{0.5}} \put(54,19){\circle*{0.5}}
  \put(55,19){\circle*{0.5}} \put(56,19){\circle*{0.5}}
  \put(57,18){\makebox(0,0)[bl]{C}}
  \put(58,13){\line( 0,1){4}}
  \put(59,13){\line( 0,1){0.8}}
  \put(59,14.6){\line( 0,1){0.8}} \put(59,16.2){\line( 0,1){0.8}}
  \put(57, 9){\makebox(0,0)[bl]{O}}
  \put(60,20.5){\line( 2,1){4}}
  \put(60,19.5){\circle*{0.5}} \put(61,20){\circle*{0.5}}
  \put(62,20.5){\circle*{0.5}} \put(63,21){\circle*{0.5}}
  \put(64,21.5){\circle*{0.5}}
  \put(65,22){\makebox(0,0)[bl]{N}}
  \put(66.5,26){\line( 0,1){4}}
  \put(65,31){\makebox(0,0)[bl]{H}}
  \put(72,20){\line(-2,1){4}}
  \put(69,20){$\circlearrowright$}
  \put(70,23){$\phi$}
  \put(73,18){\makebox(0,0)[bl]{C$_\alpha$}}
  \put(72,17){\line(-2,-1){4}}
  \put(76,17){\line( 2,-1){4}}
  \put(65,13){\makebox(0,0)[bl]{H}}
  \put(77,15){$\circlearrowright$}
  \put(76,13){$\chi$}
  \put(81,13){\makebox(0,0)[bl]{R}}
  \put(76,20){\line( 2,1){4}}
  \put(77,20){$\circlearrowright$}
  \put(76,23){$\psi$}
  \put(81,22){\makebox(0,0)[bl]{C}}
  \put(82,26){\line( 0,1){4}}
  \put(83,26){\line( 0,1){0.8}}
  \put(83,27.6){\line( 0,1){0.8}} \put(83,29.2){\line( 0,1){0.8}}
  \put(81,31){\makebox(0,0)[bl]{O}}
  \put(88,19.5){\line(-2,1){4}}
  \put(88,20.5){\circle*{0.5}} \put(87,21){\circle*{0.5}}
  \put(86,21.5){\circle*{0.5}} \put(85,22){\circle*{0.5}}
  \put(84,22.5){\circle*{0.5}}
  \put(89,18){\makebox(0,0)[bl]{N}}
  \put(90.5,13){\line( 0,1){4}}
  \put(89, 9){\makebox(0,0)[bl]{H}}
  \put(92,20){\line( 2,1){4}}
  \put(97,22){\makebox(0,0)[bl]{C$_\alpha$}}
  \put(96,25){\line(-2, 1){4}}
  \put(100,25){\line( 2, 1){4}}
  \put(89,27){\makebox(0,0)[bl]{H}}
  \put(105,27){\makebox(0,0)[bl]{R}}
  \put(104,20){\line(-2,1){4}}
  \put(105,18){\makebox(0,0)[bl]{C}}
  \put(106,13){\line( 0,1){4}}
  \put(107,13){\line( 0,1){0.8}}
  \put(107,14.6){\line( 0,1){0.8}} \put(107,16.2){\line( 0,1){0.8}}
  \put(105, 9){\makebox(0,0)[bl]{O}}
  \put(108,20.5){\line( 2,1){4}}
  \put(108,19.5){\circle*{0.5}} \put(109,20){\circle*{0.5}}
  \put(110,20.5){\circle*{0.5}} \put(111,21){\circle*{0.5}}
  \put(112,21.5){\circle*{0.5}}
  \put(113,22){\makebox(0,0)[bl]{N}}
  \put(114.5,26){\line( 0,1){4}}
  \put(113,31){\makebox(0,0)[bl]{H}}
  \put(116,23){\circle*{0.5}} \put(117,23){\circle*{0.5}}
  \put(118,23){\circle*{0.5}} \put(119,23){\circle*{0.5}}
\end{picture}
}
\vspace*{-4mm}
\caption{Chemical structures of (a) amino acid, (b) polypeptide chain.}
\vspace*{-2mm}
\label{fig:structures}
\end{figure}

The next step is to figure out the number of ways a polypeptide chain can be
folded on the diamond lattice. Along the backbones of real polypeptide chains,
all the bonds are not of the same length, all the bond angles are not exactly
equal to the tetrahedral value (in particular the peptide bond is planar),
and the nitrogen atoms do not have exactly the same properties as the carbon
atoms \cite{nitrogen}.
The folding of the polypeptide chain on the regular diamond lattice, therefore,
is necessarily approximate. But the variations of bond lengths and angles are
within 10\%, and so we can expect to see features of the diamond lattice in
the folded polypeptide chains.

The diamond lattice is a face-centred cubic lattice with a two-point basis.
It can accommodate the polypeptide chain reasonably well, with the peptide
bonds in the ``trans'' configuration. The rare ``cis'' configuration of the
peptide bond can be accommodated too, using the well-known local switch
between the face-centred cubic lattice and the hexagonal lattice. (Most of
the natural peptide bonds occur in ``trans'' configuration, but a few percent
occur in the ``cis'' configuration.) All possible configurations of the
polypeptide chain can be enumerated, by specifying for every peptide bond
the location of the next peptide bond in the chain. Each peptide unit of the
chain contains two rotatable bonds of the $C_\alpha$ atom, labeled by angles
$\phi$ and $\psi$ in Fig.1. The tetrahedral geometry gives three possible
values for each of these angles, separated by $120^\circ$ steps. There are
thus nine possible orientations of each peptide unit in the chain.

The Ramachandran map describes the allowed values of the angles $\phi$ and
$\psi$ in real polypeptide chains, after taking into account hard core
repulsion between the atoms of the polypeptide backbone \cite{ramachandran}.
As Fig.2 shows, even with all the variations in bond lengths and angles,
the discrete diamond lattice approximation is not a bad starting point for
describing the preferred orientations of the angles $\phi$ and $\psi$. (The
Ramachandran map for achiral glycine is different. Unlike the other chiral
L-type amino acids, glycine can occupy the region around $\phi=60^\circ,
\psi=-60^\circ$.) To put it differently, though nature has not been able to
perfectly implement the ideal diamond lattice geometry of structural design,
evolution has taken her impressively close to it. It is also worthwhile to
observe that in general there are many ways a folded chain can cover a
3-dimensional shape. So it is not necessary that all the nine angular
orientations are occupied equally---it is sufficient to have all the
orientations as possibilities. 

\begin{figure}[tbh]
{
\vspace*{-4mm}
\epsfxsize=10cm
\centerline{\epsfbox{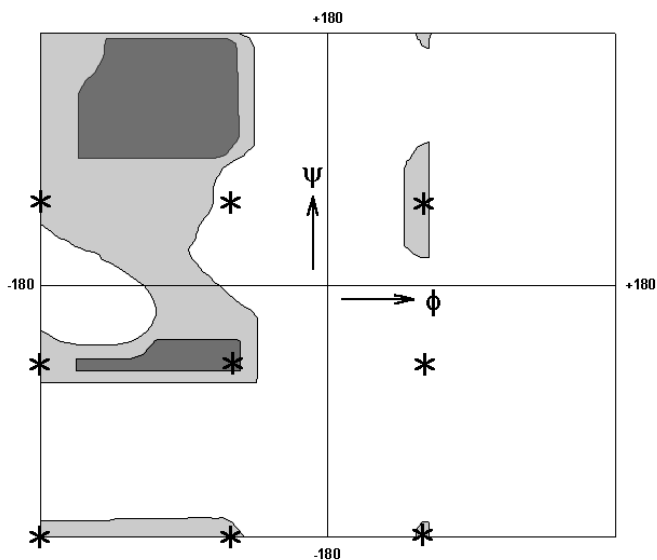}}
}
\vspace*{-4mm}
\caption{The Ramachandran map for chiral L-type amino acids, displaying the
allowed orientation angles for the $C_\alpha$ bonds in real polypeptide
chains, after taking into account hard core repulsion between all the atoms
[14]. The angles $\phi$ and $\psi$ are periodic. In the approximation that
embeds the polypeptide chain on a diamond lattice, only nine discrete
possibilities exist for the angles. These are marked by stars on the same
plot; they are uniformly separated by $120^\circ$ steps.}
\label{fig:psiphiplot}
\end{figure}

The nine angular orientations and the ``trans-cis'' option exhaust the
``elementary logic operations'' for embedding the polypeptide backbone on
the diamond lattice, i.e. by implementing these operations one can fold the
backbone on the lattice in any desired configuration. Real polypeptide chains
form other long distance connections, hydrogen and disulfide bonds, when they
fold. While these are important for structural stability of the proteins,
they do not give rise to new configurational possibilities.

\smallskip\noindent{\bf Exercise:}
{\small In two dimensions, the simplex is a triangle and the simplest
connectivity leads to a honeycomb structure. At the molecular scale,
such a geometry is also realised by carbon in the form of graphite
sheets. Figure out the building blocks required to fold a chain on the
graphite lattice in any arbitrary manner.}

\subsection{The R-groups of amino acids}

For each polypeptide backbone configuration described above, there are two
remaining directions for other chemical groups to attach to the $C_\alpha$
atom. One is attached to the R-group, while the other to a hydrogen atom.
Two arrangements are possible, and they correspond to opposite chirality.
All the amino acids naturally occurring in proteins use only one of these
two possibilities, L-type arrangement \cite{chirality}.
Altogether, therefore, there are only nine discrete orientations (and an
occasional ``trans-cis'' switch) for adding a new amino acid to an existing
polypeptide chain.

There is no clear association of any individual amino acid with the nine
discrete points in the Ramachandran map. Indeed, just the structure of a
particular amino acid does not decide its orientation in the polypeptide
chain; rather its orientation is fixed by the overall interactions of its
R-group with those that precede it and those that follow it. In other
words, the structural code of the polypeptide chain is an overlapping one.
Even with an overlapping code, every time an amino acid is added to the
polypeptide chain, the orientation of one amino acid gets decided. This
conservation of information implies that a maximally efficient overlapping
code still needs at least nine amino acids (may be 10 to take care of
the ``trans-cis'' option and long distance connections) to construct
polypeptide chains that can cover arbitrary shapes. The R-group properties
of amino acids have been studied in detail---polar and non-polar, positive
and negative charge, straight chains and rings, short and long chains, and
so on. Still which sequence of amino acids will lead to which conformation
of the polypeptide chain is an exercise, in coding as well as in chemical
properties, that has not been solved yet.

\begin{table}[tbh]
\begin{center}
\begin{tabular}{|l|l|r|r|}
\hline
Amino acid          & R-group property & Mol. wt. & Class \\
\hline
Gly (Glycine)       & Non-polar        & 75       & II \\  
Ala (Alanine)       & Non-polar        & 89       & II \\  
Pro (Proline)       & Non-polar        & 115      & II \\  
Val (Valine)        & Non-polar        & 117      & I  \\  
Leu (Leucine)       & Non-polar        & 131      & I  \\  
Ile (Isoleucine)    & Non-polar        & 131      & I  \\  
\hline
Ser (Serine)        & Polar            & 105      & II \\  
Thr (Threonine)     & Polar            & 119      & II \\  
Asn (Asparagine)    & Polar            & 132      & II \\  
Cys (Cysteine)      & Polar            & 121      & I  \\  
Met (Methionine)    & Polar            & 149      & I  \\  
Gln (Glutamine)     & Polar            & 146      & I  \\  
\hline
Asp (Aspartate)     & Negative charge  & 133      & II \\  
Glu (Glutamate)     & Negative charge  & 147      & I  \\  
\hline
Lys (Lysine)        & Positive charge  & 146      & II \\  
Arg (Arginine)      & Positive charge  & 174      & I  \\  
\hline
His (Histidine)     & Ring/Aromatic    & 155      & II \\  
Phe (Phenylalanine) & Ring/Aromatic    & 165      & II \\  
Tyr (Tyrosine)      & Ring/Aromatic    & 181      & I  \\  
Trp (Tryptophan)    & Ring/Aromatic    & 204      & I  \\  
\hline
\end{tabular}
\caption{Properties of the amino acids depend on their side chain R-groups.
Larger molecular weights indicate longer side chains. The 20 amino acids
naturally occurring in proteins have been divided into two classes of 10 each,
depending on the properties of aminoacyl-tRNA synthetases that bind the amino
acids to tRNA. These classes divide amino acids with each R-group property
equally, the longer side chains correspond to class I and the shorter ones
correspond to class II. Some specific properties not explicit in the table
are: asparagine is a shorter side chain version of glutamine, histidine has
an R-group with a small positive charge but it is close to being neutral,
and both the sulphur containing amino acids (cysteine and methionine) belong
to class I.}
\end{center}
\vspace*{-4mm}
\end{table}

At this stage, it is instructive to observe that the 20 amino acids are
divided into two classes of 10 each, according to the properties of their
aminoacyl-tRNA synthetases \cite{moras,lewin}.
tRNA molecules, with 3 nucleotide bases of anticodon at one end and an amino
acid at the other end, translate the information from mRNA molecules to
polypeptide chains. Aminoacyl-tRNA synthetases are the bilingual molecules
that ensure correct translation of this information, by attaching matching
amino acids to tRNA molecules as per their anticodons. The two classes of
synthetases totally differ from each other in their active sites and in how
they attach amino acids to the tRNA molecules. The lack of any apparent
relationship between the two classes of synthetases has led to the conjecture
that the two classes evolved independently, and early forms of life could
have existed with proteins made up of only 10 amino acids of one type or the
other. An inspection of the R-group properties of amino acids in the two
classes reveals that each property (polar, non-polar, ring/aromatic, positive
and negative charge) is equally divided amongst the two classes, as shown in
Table 4. Not only that, the heavier amino acids with each property belong to
class I, while the lighter ones belong to class II. This clear binary
partition of the amino acids according to the size of their side chains
has unambiguous structural significance. The diamond lattice structure is
quite loosely packed with many cavities of different sizes. The use of long
R-groups to fill up big cavities and short R-groups to fill up small ones
can produce a dense compact structure. Proteins indeed have a packing fraction
similar to closest packing of identical spheres.

We have thus arrived at a structural explanation for the 20 amino acids as
building blocks of proteins, a factor of 10 for folding the polypeptide
backbone and a factor of 2 for the size of the R-group \cite{oddones}.
This explanation can be made more concrete by identifying subsequences of
amino acids corresponding to different orientations of the polypeptide
backbone. We do not have a clear criterion regarding how long the amino
acid sequence should be before it assumes a definite orientation, although
the short range nature of molecular interactions favours relatively local
folding rules. The information accumulated in protein structure databases
should help in a detailed analysis.

\smallskip\noindent{\bf Exercise:}
{\small Devise a scheme to discretise the angular distributions of amino
acids on the Ramachandran map. Check how the distributions narrow down,
when preceding and following amino acids are fixed, step by step.}

\section{Language of DNA}

Now we can analyse the other task of synthesising linear chains by joining
together the building blocks \cite{quant_gc}.
This task needs an external agency---called an oracle in the language of
computer science---to dictate the required order of building blocks. DNA
and polypeptide chains are assembled in presence of preexisting templates;
the order of building blocks specified by the template is the blueprint or
the master copy.

To understand the optimisation of this process, it is helpful to first look
at a familiar game played by school children. In the game, two teams compete
to find the names of famous persons. One team selects the name of a famous
person and gives it to the referee. The other team has to discover this name
by asking a set of questions to the first team. The game is made interesting
by the condition that the first team only provides ``yes or no'' answers to
the questions, i.e. the minimal amount of information---one bit---is released
in response to every question. The two teams take turns choosing the names
and asking the questions. After several rounds, the team which succeeds in
discovering the names with a smaller number of questions is the winner.

The players quickly learn that specific questions such as ``Is the person
Albert Einstein?'' are inefficient. They fail most of the time, and when they
fail one does not learn much about who the person is. Efficient questions
are of the type ``Is the person a man or a woman?'', whence there is a
substantial reduction in the number of possible names for the next step no
matter what the answer to the question is. The best questions are the ones
which reduce the number of possible names for the next step by a factor of
two. This factor of two reduction is of course a consequence of the fact
that the answers provided to the questions are binary.

The computer science paradigm for this game is ``database search''. Binary
search is the optimal classical algorithm, which finds the desired item in
the database using $\log_2 N$ binary questions. This algorithm can be used
when the items in the database are sorted according to some order, so that
at every step the items corresponding to ``yes'' answer can be easily
separated from the items corresponding to ``no''. Examples are words in a
dictionary and names in a telephone directory. On the other hand, if the
database is not sorted, e.g. numbers in a telephone directory, then there
is no fast search algorithm. The best option is to go through all the items
one by one, and that requires $N/2$ binary questions on the average.

Now we can return to look at how the DNA and polypeptide chains are assembled
by biochemical processes. The questions are the molecular bonds involved in
nucleotide base-pairing, the preexisting template is the ``other team'', and
the answers are binary---either the bonds form or they do not. The available
database is unsorted---the building blocks are randomly floating around in
the cellular environment. Classically, two nucleotide bases (one complementary
pair) would be sufficient to encode genetic information; the number of more
complicated building blocks would then follow powers of two. Our computers
indeed encode information this way. Moreover, starting from scratch, two
nucleotide bases would have evolved before four nucleotide bases. So why did
living organisms give up the simple powers of two and evolve more complicated
languages?

A surprising answer is provided by the quantum database search algorithm,
whose optimisation features are different from their classical analogues.

\subsection{Quantum database search}

Quantum states are unit vectors in a Hilbert space (linear vector space with
complex coefficients). They evolve in time by unitary transformations. These
properties are totally different from classical states and their evolution,
and they form the basis of the subject of quantum computation. Of course, to
make contact with our problems defined in classical language, we require a
mapping between classical and quantum states. That is achieved by identifying
the orthogonal basis vectors of the Hilbert space with the set of distinct
classical states. The complex components of a general vector in the Hilbert
space can vary continuously, and they are called amplitudes of the quantum
state. States with more than one non-zero amplitudes are called superposition
states. Quantum algorithms evolve the amplitudes from some initial values to
some final values, by a sequence of unitary transformations. The measurement
probability of obtaining a particular classical result from a quantum state
is given by the absolute value square of the corresponding amplitude.

The quantum database search algorithm works in an $N$-dimensional Hilbert
space, whose basis vectors are identified with individual items. It takes
an initial state whose amplitudes are uniformly distributed over all items,
to a final state where all but one amplitudes vanish. Following Dirac's
notation, let $\{|i\rangle\}$ be the set of basis vectors, $|s\rangle$ the
initial uniform superposition state, and $|b\rangle$ the final state
corresponding to the desired item. Then
\begin{equation}
|\langle i|s \rangle| = 1/\sqrt{N} ~,~~ \langle i|b \rangle = \delta_{ib} ~~.
\end{equation}
Since superposition of amplitudes is commutative, the quantum database can
be taken to be unsorted without any loss of generality. The optimal solution
to the quantum search problem \cite{grover}
is based on two properties: (i) the shortest path between two points on the
unitary sphere is the geodesic great circle, and (ii) the largest step one
can take in a given direction, consistent with unitarity, is the reflection
operation. The available reflection operators are,
\begin{equation}
U_b = 1 - 2|b\rangle\langle b| ~,~~ -U_s = 2|s\rangle\langle s| - 1 ~~.
\end{equation}
The former is the binary quantum query for the desired state, while the latter
is the sign flip relative to the initial uniform state. The quantum algorithm
is the discrete Trotter's formula with these two operators, which locates the
desired item in the database with $Q$ queries,
\begin{equation}
(-U_sU_b)^Q |s\rangle = |b\rangle ~~.
\end{equation}
This equation is readily solved to give the relation
\begin{equation}
(2Q+1) \sin^{-1} (1/\sqrt{N}) = \pi/2 ~~.
\end{equation}
For a given $N$, the solution for $Q$ satisfying Eq.(6) may not be an integer.
This means that the algorithm will have to stop without the final state being
exactly $|b\rangle$. There will remain a small admixture of other amplitudes
in the output, implying an error in the search process. The size of the
unwanted admixture is determined by how close one can get to $\pi/2$ on the
r.h.s. of Eq.(6). Apart from this, the algorithm is fully deterministic.

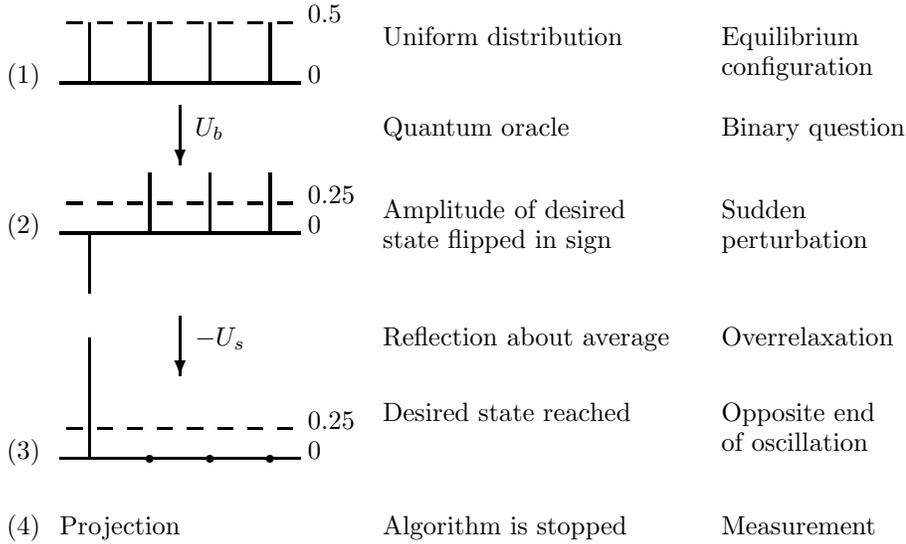
\begin{figure}[tbh]
{
\setlength{\unitlength}{1mm}
\begin{picture}(120,75)
  \thicklines
  \put( 5,65){\makebox(0,0)[bl]{(1)}}
  \put(12,65){\line(1,0){32}}
\put(13,73){\line(1,0){2}} \put(17,73){\line(1,0){2}} \put(21,73){\line(1,0){2}}
\put(25,73){\line(1,0){2}} \put(29,73){\line(1,0){2}} \put(33,73){\line(1,0){2}}
\put(37,73){\line(1,0){2}} \put(41,73){\line(1,0){2}}
  \put(45,65){\makebox(0,0)[bl]{0}} \put(45,73){\makebox(0,0)[bl]{0.5}}
  \put(16,65){\line(0,1){8}} \put(24,65){\line(0,1){8}}
  \put(32,65){\line(0,1){8}} \put(40,65){\line(0,1){8}}
  \put(55,70){\makebox(0,0)[bl]{Uniform distribution}}
  \put(100,70){\makebox(0,0)[bl]{Equilibrium}}
  \put(100,66){\makebox(0,0)[bl]{configuration}}
  \put(28,62){\vector(0,-1){8}}
  \put(30,58){\makebox(0,0)[bl]{$U_b$}}
  \put(55,58){\makebox(0,0)[bl]{Quantum oracle}}
  \put(100,58){\makebox(0,0)[bl]{Binary question}}
  \put( 5,45){\makebox(0,0)[bl]{(2)}}
  \put(12,45){\line(1,0){32}}
\put(13,49){\line(1,0){2}} \put(17,49){\line(1,0){2}} \put(21,49){\line(1,0){2}}
\put(25,49){\line(1,0){2}} \put(29,49){\line(1,0){2}} \put(33,49){\line(1,0){2}}
\put(37,49){\line(1,0){2}} \put(41,49){\line(1,0){2}}
  \put(45,45){\makebox(0,0)[bl]{0}} \put(45,49){\makebox(0,0)[bl]{0.25}}
  \put(16,45){\line(0,-1){8}} \put(24,45){\line(0,1){8}}
  \put(32,45){\line(0,1){8}} \put(40,45){\line(0,1){8}}
  \put(55,47){\makebox(0,0)[bl]{Amplitude of desired}}
  \put(55,43){\makebox(0,0)[bl]{state flipped in sign}}
  \put(100,47){\makebox(0,0)[bl]{Sudden}}
  \put(100,43){\makebox(0,0)[bl]{perturbation}}
  \put(28,34){\vector(0,-1){8}}
  \put(30,30){\makebox(0,0)[bl]{$-U_s$}}
  \put(55,30){\makebox(0,0)[bl]{Reflection about average}}
  \put(100,30){\makebox(0,0)[bl]{Overrelaxation}}
  \put( 5,15){\makebox(0,0)[bl]{(3)}}
  \put(12,15){\line(1,0){32}}
\put(13,19){\line(1,0){2}} \put(17,19){\line(1,0){2}} \put(21,19){\line(1,0){2}}
\put(25,19){\line(1,0){2}} \put(29,19){\line(1,0){2}} \put(33,19){\line(1,0){2}}
\put(37,19){\line(1,0){2}} \put(41,19){\line(1,0){2}}
  \put(45,15){\makebox(0,0)[bl]{0}} \put(45,19){\makebox(0,0)[bl]{0.25}}
  \put(16,15){\line(0,1){16}}
  \put(24,15){\circle*{1}} \put(32,15){\circle*{1}} \put(40,15){\circle*{1}}
  \put(55,20){\makebox(0,0)[bl]{Desired state reached}}
  \put(100,20){\makebox(0,0)[bl]{Opposite end}}
  \put(100,16){\makebox(0,0)[bl]{of oscillation}}
  \put( 5,5){\makebox(0,0)[bl]{(4)}}
  \put(12,5){\makebox(0,0)[bl]{Projection}}
  \put(55,5){\makebox(0,0)[bl]{Algorithm is stopped}}
  \put(100,5){\makebox(0,0)[bl]{Measurement}}
\end{picture}
}
\vspace*{-2mm}
\caption{The steps of the quantum database search algorithm for the simplest
case of 4 items, when the first item is desired by the oracle. The left column
depicts the amplitudes of the 4 states, with the dashed lines showing their
average values. The middle column describes the algorithmic steps, and the
right column mentions their physical implementation.}
\vspace*{-4mm}
\label{fig:database}
\end{figure}

\noindent{\bf Exercise:}
{\small Derive the result in Eq.(6). Show that it is the optimal solution.}
\smallskip

For unsorted database search, the result in Eq.(6) is clearly superior to
the corresponding classical result (the number of queries is asymptotically
$O(\sqrt{N})$ vs. $O(N)$). The reason behind this improvement is the clever
interference amongst superposed amplitudes, which amplifies the desired
state amplitude at the expense of the rest. As a matter of fact, this quantum
algorithm does not use the full power of quantum dynamics. It can very well
be implemented using other systems that obey the superposition principle,
e.g. classical waves or coupled pendulums \cite{wavesearch}.
Fig.3 illustrates the individual steps of the algorithm in the simplest case.

The solutions of Eq.(6) for small values of $Q$ are of special significance
for the number of building blocks involved in genetic information processing,
\begin{eqnarray}
Q=1 &\Rightarrow& N=4    ~, \nonumber\\
Q=2 &\Rightarrow& N=10.5 ~, \nonumber\\
Q=3 &\Rightarrow& N=20.2 ~,
\end{eqnarray}
when the binary quantum query is identified with the nucleotide base-pairing.
DNA has an alphabet of 4 nucleotide bases, the well-known triplet genetic
code for synthesis of polypeptide chains carries 21 signals (for 20 amino
acids plus STOP) \cite{watson,lewin},
and a primitive doublet genetic code could have labeled the 10 amino acids
of a single class (described in the previous section). It is fascinating
that these numbers have come out of an algorithm that performs the actual
task accomplished by DNA \cite{altformula},
and that too as optimal solutions \cite{errors}.
Obviously, the next job is to check whether DNA possesses the hardware
components necessary to implement the quantum search algorithm.

\begin{figure}[tbh]
\epsfxsize=7.6cm
\centerline{\epsfbox{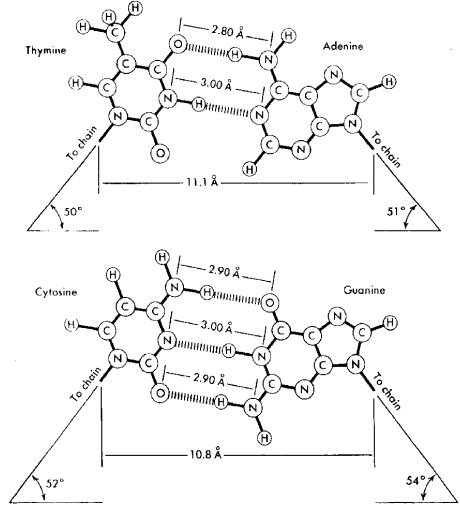}}
\caption{The nucleotide base-pairing in DNA with hydrogen bonds [22].}
\vspace*{-4mm}
\end{figure}

\subsection{A physical scenario}

The execution of the quantum search algorithm requires three ingredients:
an interaction providing the quantum oracle $U_b$, a mechanism to generate
the superposition state $|s\rangle$, and an environment where quantum
coherence can survive. Let us look at how these may be realised in the case
of DNA.

The formation of a molecular bond is a quantum process. Two reactants come
together in an excited state, and the combined system decays to the bound
state with the emission of an energy quantum. The total energy is conserved
in the process, but the quantum dynamics of the transition between the two
molecular energy levels alters the phase of the quantum state by $\sqrt{-1}$.
This phase is a geometric phase, independent of the strength of the transition
interaction and the time for the decay.

The DNA molecule is a double helix, and is often schematically represented
as a ladder. The sides of the ladder have a periodic structure made of
alternating sugar and phosphate groups, while the rungs of the ladder are
made of paired nucleotide bases. The atomic structure of paired nucleotide
bases is shown in Fig.4. The nucleotide bases are made of planar rings of
atoms, and so the base-pairing requires at least two points of contact for
structural stability of the helix (otherwise the bases would rotate relative
to each other). The contacts between nucleotide bases are hydrogen bonds,
which are formed by the quantum process of a proton tunneling between two
attractive energy minima. When the nucleotide bases come together with
arbitrary initial orientations, the base-pairing is likely to occur as a
two step process---the first contact connects the two nucleotide bases and
the second one locks their relative orientations. With such a two step
base-pairing, the quantum state is transformed precisely by the operator
$U_b$, i.e. the quantum state changes its sign when the nucleotide bases get
paired and it remains unaffected when there is no pairing. From the known
binding energy of the hydrogen bond and the uncertainty principle, the time
scale for base-pairing can be estimated to be, $t_b \sim 10^{-14}$ sec.

\begin{figure}[tbh]
\epsfxsize=8cm
\centerline{\epsfbox{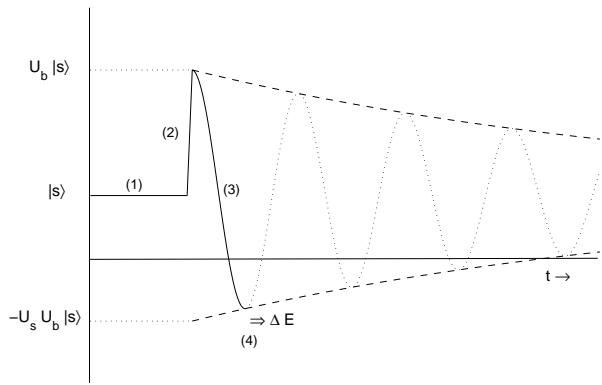}}
\vspace*{-2mm}
\caption{A time evolution scenario for the DNA replication process.
Numerical labels refer to the steps described in Fig.3.}
\vspace*{-4mm}
\end{figure}

The quantum algorithm starts with the superposition state $|s\rangle$, so
$|s\rangle$ has to be a stable equilibrium state. Any other initial state
has to relax towards $|s\rangle$, and let $t_r$ be the time scale for this
relaxation process. In case of DNA, $|s\rangle$ is a superposition state
of chemically distinct nucleotide bases; it can be created only if the
environment provides transition matrix elements between its various components.
The nucleotide bases differ from each other by 5-10 atoms, and the transition
matrix elements between them vanish in free space. Exchange of atomic groups
is routine in chemical reactions, however---these reactions take place not
with fixed number of atoms but with fixed chemical potential where exchange of
atoms with the environment is allowed. (It is important to note that chemical
reactions produce mixtures in classical environments while they produce
superpositions in quantum environments.) The magnitudes of the transition
matrix elements decide the speed at which $|s\rangle$ cycles through its
various components. Let $t_{osc}$ be the time scale for these oscillations.

As illustrated in Fig.5, the quantum search algorithm can be executed when
the above defined time scales satisfy the hierarchy,
\begin{equation}
t_b << t_{osc} << t_r ~.
\end{equation}
Explicitly, (1) the environment relaxes the initial state to $|s\rangle$,
(2) the sudden base-pairing changes the state to $U_b|s\rangle$ (the energy
quantum is not released yet), (3) the relaxation process tries to bring the
state back to $|s\rangle$ causing damped oscillations, and (4) the energy
quantum becomes free to wander off at the opposite end of oscillation; its
irreversible release confirms the base-pairing and stops the search algorithm,
$-U_sU_b|s\rangle = |b\rangle$.

This scenario is quite robust, and does not rely on fine-tuned parameters.
The all important remaining question is whether the environment of DNA really
has the properties ascribed to it, i.e. can it protect quantum coherence for
a long enough time and create superposition states?

\subsection{Decoherence and enzymes}

High precision experiments have shown that quantum dynamics is the only valid
description of the happenings at the atomic scale. Nevertheless, we hardly
observe quantum effects in macroscopic systems; we find that macroscopic
systems follow the quite different rules of classical dynamics. It cannot be
discounted that the cross-over in the behaviour of physical systems is caused
by some new laws of physics, as yet undiscovered. On the other hand, by a
careful analysis of quantum dynamics, it is possible to explain the cross-over
in a self-consistent manner. This explanation is labeled decoherence. It is
a consequence of the fact that no quantum system is perfectly isolated; its
inevitable interactions with its environment lead to irreversible loss of
information and classical behaviour \cite{zeh}.
Irrespective of any undiscovered laws of physics, therefore, decoherence must
be controlled in order to observe quantum effects (e.g. in quantum computers).

It is impossible to give a complete quantitative description of decoherence,
because sufficient details of the environment are not available in general. 
Still we can construct a formulation good enough to provide order-of-magnitude
estimates. The dominant interactions of a system with its environment are
molecular collisions and long range forces. The cross-sections for these
processes can be calculated in terms of global properties of the environment,
such as temperature and density. Then the rate at which the system relaxes
towards the equilibrium state favoured by the environment is estimated
according to Fermi's golden rule. (This relaxation is often referred to as the
``collapse of the quantum wavefunction''.) The relaxation rate is inversely
proportional to three factors: the initial flux, the interaction strength and
the final density of states. Typical decoherence times are extremely short,
beyond direct observation. But we also know specific situations, where quantum
states are long-lived due to suppression of one or more of these factors.
For instance, low temperatures and shielding reduce the initial flux, the
interaction strength is small for lasers and nuclear spins, and the final
density of states is suppressed for superconductors and hydrogen bonds due
to energy gap.

It must be noted that the relaxation rate cannot become arbitrarily small
when these factors become large. It is well-known that classical waves cannot
be damped faster than the critical rate of damping, given by their natural
period of oscillation. If the damping is increased beyond the critical value,
the system stops relaxing instead of relaxing faster. The quantum analogue of
this is the Zeno effect, which points out that a continuously monitored system
cannot decay. The reason behind this caveat is that Fermi's golden rule is an
approximation, not valid at times smaller than the natural oscillation period.
It follows that, in the notation of the previous subsection, $t_{osc} \leq t_r$
\cite{critdamp}.

Biochemical reactions take place in a liquid medium at room temperature.
In this environment, the decoherence times estimated following conventional
thermodynamics and Fermi's golden rule are miniscule. On the other hand,
it is known that highly specific enzymes enhance the rates of biochemical
reactions by orders of magnitude (as large as $10^{12}$) compared to estimates
of kinetic theory and diffusion processes. Enzymes are catalysts, and the
standard explanation of their properties is that they speed up the reactions
by binding to the transition states and lowering the reaction barriers
\cite{lehninger}.
What the transition states are and how the reaction barriers are lowered must
be explained ultimately in terms of underlying physical laws. A transition
state is something in between the reactants and the products, which can be
represented by distorted quantum wavefunctions, i.e. a superposition. The
intermediate transition state will not be stable on its own, but the enzymes
are able to store and supply free energy necessary for stabilising it.
The reduction of reaction barrier by stabilisation of superposition states
can provide a large speed up, because the tunneling amplitudes depend
exponentially on the barrier size. Thus it is entirely possible that the
enzymes perform their job by exploiting the quantum superposition phenomenon.

DNA replication takes place only in the presence of polymerase enzymes,
so we must focus on the kind of environment they provide. The polymerase
enzymes are much larger than the nucleotide bases, and they completely
enclose the reaction region during replication. This behaviour automatically
shields the reaction from the flux of external disturbances and reduces the
final density of states by limiting possible configurations (the allowed
phase space is essentially one-dimensional and not three-dimensional). The
hydrogen bonds also reduce the final density of states, since their binding
energy ($\sim 7 kT$) is considerably larger than the thermal fluctuations.
The polymerase enzymes drive the replication process monotonically, using
free energy from ATP molecules, which is another indication that the thermal
fluctuations are overcome.

Altogether, it is not inconceivable for a cleverly designed polymerase enzyme
to create superposition states and protect quantum coherence in accordance
with Eq.(8). Moreover, certain aspects of the quantum replication algorithm
can be tested experimentally: artificial DNA with varying number of nucleotide
bases can show whether 4 is the optimal number of nucleotide bases or not,
and isotopic tagging can track whether superposition of atoms takes place
or not \cite{testdna}.

\section{Outlook}

Information theory provides a powerful framework for extracting essential
features of complicated processes of life, and then analysing them in a
systematic manner. The easiest processes to study are no doubt the ones at
the lowest level. We have learnt a lot, both in computer science and in
molecular biology, since their early days \cite{schrodinger,neumann,frozen},
and so we can now perform a much more detailed analysis. Physical theories
often start out as effective theories, where the predictions of the theories
depend on certain parameters. The values of the parameters have to be either
assumed or taken from experiments; the effective theory cannot predict them.
To understand why the parameters have the values they do, we have to go one
level deeper---typically to smaller scales, e.g. viscosity of fluids and
crystal shapes of solids can be understood based on atomic forces, structure
of atoms can be understood based on properties of electrons and quarks, and
so on. When the deeper level reduces the number of unknown parameters, we
consider the theory to be more complete and satisfactory. The level below
conventional molecular biology is spanned by atomic structure and quantum
dynamics, and that is the natural place to look for reasons behind life's
``frozen accident''. It is indeed wonderful that sufficient ingredients exist
at this deeper level to explain the frozen accident as the optimal solution.

Counting the number of building blocks in the languages of DNA and proteins
is only the first step. The obvious next step to investigate is why the
languages use only the observed building blocks and not other similar ones;
only a subset of nucleotide bases and amino acids existing in living cells
are used as the building blocks. The likely criteria for the selection of
particular building blocks are simplicity (for easy availability and quick
synthesis) and functional ability (for implementing the desired tasks). In
case of proteins, the choices are connected to the solution of the protein
folding problem, i.e. which way a particular amino acid sequence will fold.
In case of DNA, the choices are possibly connected with the magnitudes of
the transition matrix elements. We do not have the answers yet, but we may
not have to wait very long. The experimental techniques and the information
collected in databases have reached a stage, where it is possible to form
hypotheses and then test them. It would certainly be worthwhile to test the
arguments presented here in detail, and then build on them to understand
more and more complicated processes of life.

The results described here also allow us to speculate about the origin of
life. All the complicated processes of life did not come about in one go.
The lowest level of information processing not only arose first, but in
all likelihood it was much simpler than what exists today. Importance of
functionality over memory would put proteins before DNA.  Similar codons
for similar amino acids and wobble rules found in the present genetic code
are possible relics of an earlier simpler system \cite{wobble}.
The solutions of the quantum search algorithm and the amino acid classes
suggest that the present triplet genetic code was preceded by a doublet one.
(This idea is reinforced by the accidental degeneracy, where 20 amino acids
can be coded either by one classical and two quantum queries or by three
quantum queries.) A still simpler precursor to the doublet code would be a
singlet code of 4 amino acids. In fact such an evolutionary route for the
genetic code has been proposed, just based on the chemical properties of
the amino acids, GNC $\rightarrow$ SNS $\rightarrow$ NNN \cite{ikehara}.

Unraveling the mysteries of life is certainly exciting. It is not merely a
theoretical adventure; it has immense potential applications. Medical science
will obviously benefit if we learn how to control processes of life that go
wrong. Nanotechnology would benefit as well if we learn from biology how to
implement specific tasks at the molecular scale.

\section*{Acknowledgements}
Many persons, too numerous to name individually, have helped in this study
by guiding my thoughts in various directions. I am grateful to the Center
for Computational Physics, University of Tsukuba, Japan, and the Optical
Physics Research group, Lucent Technologies, USA, for their hospitality
during part of this work.


\begin{thebibliography}{99}
\bibitem{levels}{Compared to deciphering an undocumented compiler written
         by someone else, it is much easier to write one's own compiler
         from scratch, as anyone who has struggled with such problems would
         testify.}
\bibitem{norm}{``Orthogonal transformations'' have been enclosed in quotes
         here because it is the linear norm of the vector that is conserved.
         In conventional physical applications, orthogonal transformations
         represent conservation of the quadratic norm.}
\bibitem{zerosum}{The general subject of ``zero-sum games'' has not been
         explored in detail. The same arguments can be applied in a variety
         of situations, and may run counter to ethical principles, e.g.
         capitalism will conquer socialism.}
\bibitem{bounds}{We don't need exact value of $\pi$ in practical applications
         of geometry. We can first choose the desired accuracy of the result,
         and then find an appropriate digital approximation to $\pi$ to get
         the result within those bounds.}
\bibitem{english}{Written English is non-phonetic and does not make efficient
         use of its alphabet (e.g. ``q'' is always followed by ``u'', even
         when new words like ``qubit'' are invented). But it will be a while
         before we come up with a more efficient alternative.}
\bibitem{cartesian}{The position of a point in a multi-dimensional space is
         often specified using orthogonal Cartesian coordinate system, i.e.
         a direct product of 1-dimensional coordinates. While there is nothing
         wrong with such a description, it is not the simplest choice.}
\bibitem{nitrogen}{The nitrogen atom of the peptide bond carries a positive
         charge, however, which makes its electronic behaviour similar to the
         tetravalent carbon atom.}
\bibitem{chirality}{The smallest amino acid (glycine) is achiral, with
         hydrogen atom as the R-group. All the others are naturally found
         in the chiral L-type configurations, with the smallest among them
         (alanine) only rarely displaying the D-type configuration. Detailed
         model building studies have shown that all the R-groups in a
         polypeptide chain must be of the same chirality for the stability of
         regular secondary structures ($\alpha$-helices and $\beta$-sheets).}
\bibitem{oddones}{There is one special amino acid in each class, involved in
         transformations beyond folding the backbone. Cysteine of class I
         forms long distance disulfide bonds, and proline of class I helps
         in ``trans-cis'' switch.}
\bibitem{errors}{The $Q=1$ solution is exact, and DNA replication is known
         to be quite error-free. The solutions for $Q=2,3$ are not exact.
         They correspond to an error rate of about 1 part in 1000, which is
         not in conflict with the observed error rate in protein synthesis.}
\end{thebibliography}
\end{document}